 \documentclass[aps,pra,showpacs,twocolumn,reprint,floats,epsfig,pdflatex]{revtex4-1}
\usepackage{graphicx,color}
\usepackage{url,hyperref}
\usepackage{amsmath}
\usepackage{epsfig}
\usepackage{changebar}
\usepackage{subcaption}
\usepackage{footmisc}
\usepackage{epsfig}
\usepackage{amsfonts}
\usepackage{amssymb}
\usepackage{amsbsy}
\usepackage{xcolor}
 \usepackage{braket}

\begin{document}

\title{SeD Radical: A probe for measurement of time variation of Fine Structure Constant($\alpha$) and Proton to Electron Mass Ratio($\mu$) }
\author{Gaurab Ganguly$^1$, Avijit Sen$^1$, Manas Mukherjee$^2$\footnote{author for correspondence: phymukhe@nus.edu.sg} and Ankan Paul$^1$\footnote{author for correspondence: rcap@iacs.res.in}}

\affiliation{$^1$Raman Center for Atomic, Molecular and Optical Sciences, Indian Association for the Cultivation of Science, Kolkata, INDIA}
\affiliation{$^2$ Centre for Quantum Technologies, National University Singapore, Singapore}.

\begin{abstract}
 Based on the spectroscopic constants derived from highly accurate potential energy surfaces, the SeD radical is identified as a
 spectroscopic probe for measuring spatial and temporal variation of fundamental physical constants such as the fine-structure constant
(denoted as  $\alpha=\frac{e^2}{\hbar c}$) and the proton-to-electron mass ratio (denoted as $\mu=\frac{m_p}{m_e}$). The ground state of
SeD ($X^2\Pi$), due to spin-orbit coupling, splits into two fine structure multiplets $^2\Pi_{\frac{3}{2}}$ and $^2\Pi_{\frac{1}{2}}$.
The potential energy surfaces of these spin-orbit components are derived from a state of the art electronic structure method, MRCI+Q
inclusive of scalar relativistic effects with the spin-orbit effects accounted through the Breit-Pauli operator. The relevant spectroscopic
data are evaluated using Murrel-Sorbie fit to the potential energy surfaces. The spin-orbit splitting ($\omega_f$) between the two multiplets
is similar in magnitude with the harmonic frequency ($\omega_e$)
of the diatomic molecule. The amplification factor (K) derived from this
theoretical method for this particular molecule can be as large as 350, on the lower side it can be about 34.
The significantly large values of K indicate that SeD radical can be a plaussible experimental candidate for measuring variation
in $\alpha$ and $\mu$.

\end{abstract}


\maketitle

\section{Introduction}
Spatial and temporal variation of the fundamental constants to some level
points to the invalidation of the Einstein's equivalence principle\cite{arx_GA,arx_CO}.
These include the coupling constant of electromagnetic interaction,
usually called as fine-structure constant (denoted as $\alpha=\frac{e^2}{\hbar c}$)
and the proton-to-electron mass ratio (denoted as $\mu=\frac{m_p}{m_e})$.
Over vast space and large time scale of the expanding universe these variations
can be astrophysically measured and compared to the high precision laboratory data\cite{uzan_rmp}.
Recent advances in computational methods and high-precision experimental techniques
propelled researchers to propose different experiments as well as experimental
candidates aimed at the determination of such variation in the last few decades\cite{flam_ijmp,flam_epjs}.
High precision trapped atom and molecular ion spectroscopy is one of the most promising approaches for measuring
such space-time variation of the fundamental physical constants experimentally,
as molecular spectroscopy is sensitive to both the dimensionless constants $\alpha$ and $\mu$\cite{flam_njp,flam_science}.
These spectroscopic techniques employ a diatomic molecule as a probe for measuring such
variations in fundamental physical constants following the proposal put forward by Flambaum and Kozlov\cite{flam_prl}.

Diatomic molecules or radical having nearly degenerate long lived rotational
and vibrational levels belonging to different electronic states are particularly
sensitive to measure the variation in $\alpha$ and $\mu$ due to several orders
of magnitude enhancement\cite{flam_prl}. In case of a neutral or charged diatomic molecule
having unpaired electrons with ground state fine structure multiplet
the transition frequency between the two multiplet states is given by,
\begin{equation}
\omega=\omega_f - v \omega_e  \; ; \;\;\;\;\;\;  v=1,2,3,4,..
\end{equation}

where $\omega_f$ is the magnitude of spacing between the multiplets (spin-orbit),
$\omega_e$ is the magnitude of vibrational spacing under harmonic approximation and
$v$ is the vibrational quantum number\cite{flam_prl}. The fine structure interval
$\omega_f$ holds the relation with $\alpha$ as $\omega_f\sim Z^2 \alpha^2 E_H$,
where Z is the nuclear charge and $E_H=\frac{m_e e^4}{h^2}$.
On the other hand $\omega_e$ is related to $\mu$ which is given by $\omega_e\sim M_r^{-\frac{1}{2}}\mu^{-\frac{1}{2}}E_H $.
Therefore $\omega$ is sensitive to the variation of both $\alpha$ and $\mu$ as given by the following equations,
\begin{equation}
\delta \omega = 2 \omega_f \frac{\delta \alpha}{\alpha}+\frac{v}{2}\omega_e \frac{\delta \mu}{\mu} \label{delw}
\end{equation}
On the other hand the fractional variation of $\omega$ may be written as\cite{flam_prl}

\begin{eqnarray}
\frac{\delta \omega}{\omega} &=& \frac{1}{\omega}(2 \omega_f \frac{\delta \alpha}{\alpha}
                +\frac{v}{2}\omega_e \frac{\delta \mu}{\mu}) , \nonumber \\
       %&=& K (2 \frac{\delta \alpha}{\alpha} + \frac{1}{2} \frac{\delta \mu}{\mu}) \nonumber \\
       &=& 2K (\frac{\delta \alpha}{\alpha} + \frac{1}{4} \frac{\delta \mu}{\mu}) \label{fracw},
\end{eqnarray}

where $K=\frac{\omega_f}{\omega_f - v \omega_e}=\frac{\omega_f}{\omega}$\cite{flam_prl} is known
as the enhancement factor or the amplification factor. According to Flambaum and Kozlov large value for K of a species hints at its
potential candidature as an experimental probe to gauge the variation in $\alpha$ and $\mu$.
Ideally diatomic molecules for which $\omega = 0 $ would be the best possible probes for such purpose.
However it turns out that such a possibility is purely fortuitous as no such diatomic molecules
exist. This limits the search for such diatomic molecules to cases where $\frac{\omega_f}{\omega}$ and $K$ is substantially large .
Unfortunately there are very few molecules which obey this criterion.
Therefore it is essential to identify molecular candidates on which both experiments can
be performed and astrophysical observations can be made.
Flambaum {\it et.al.} and others have recently proposed certain candidates as viable probes,
such as {Cs}$_2$\cite{demil_prl}, MgH, CaH$^+$\cite{kajita_pra,kajita_jpb}, Cl$_2^+$, IrC, HfF$^+$\cite{skripnikov_jept-lett,barker_jcp},
NH$^+$\cite{belay_pra,kawaguchi_pra,hubers_jcp}, SiBr\cite{beloy_pra}. Out of these few do not have permanent
dipole moments and are inactive to microwave spectroscopy.
Although SiBr is relevant but its harmonic stretching frequency falls out of the infrared
spectroscopy window. It is imperative to identify new candidates as this would enrich
the gamut of probes so that more systematic analysis can be conducted. In this paper
we identify Selenium Deuteride (SeD) as a potential candidate for experimentally probing
the variation of fundamental physical constants and perform detailed theoretical
study on that particular molecule to find variation in transition frequency upon
a given change in $\alpha$ and $\mu$.
Although $NH_3$ and other C-H compounds are astrophysically abundant, SeD radical
is yet to be observed. However, Asymptotic Giant Branch stars are probable sites
to look for SeD due to the abundance of s-process isotopes and freely available
deuterium in the cooler portion of the universe. The added advantage would be the cooler temperature for good IR observation
which can be potentially important complementarity to the microwave observations
made on $NH_3$ inversion. Laboratory experiment wise SeD is very similar to the
trapping of NH or CaH molecules in magneto-optical traps\cite{arx_CO,campbell_prl}
and hence it may be considered as a possible candidate for fundamental test.
Chemically, SeD is an open shell molecule with one unpaired electron $(S=\frac{1}{2})$
in its $\pi$ orbital with a $\pi^3$ configuration. The first excited state
$A^2\Sigma^+$ is well separated (about $30,460 cm^{-1}$) with the ground state
$X^2\Pi$. For linear molecule under spin-orbit coupling (SOC) the electronic states can be expressed as
$\Omega=|\Lambda+\Sigma|$, where, $\Lambda$ and $\Sigma$ are the orbital and spin angular momentum.
Under SOC splitting the $X^2\Pi$ state will split into $^2\Pi_{\frac{3}{2}}$ and $^2\Pi_{\frac{1}{2}}$
for $\Lambda=1,\Sigma= \pm \frac{1}{2}$ and there is no split for $A^2\Sigma^+$
because of $\Lambda=0,\Sigma= \frac{1}{2}$. According to the Hund's rule for more than half filled
($\pi^3$ electronic configuration) $^2\Pi_{\frac{3}{2}}$ is energetically lower than
$^2\Pi_{\frac{1}{2}}$. The fine structure and vibrational spacings of the
$X^2\Pi$ state are similar in magnitude ($\omega_f \approx v\omega_e,v=1$)\cite{jmbrown_phy_scrip}.

\section{RO-VIBRONIC ENERGY LEVELS IN Selenium Deuteride}

The total Hamiltonian can be expressed as,
\begin{eqnarray}
H = H_{Vib}+H_{SO}+H_{Rot} \label{H_tot}
\end{eqnarray}
for a $^2\Pi$ state the terms on the right hand side of the equation
represents vibronic Hamiltonian, spin-orbit interaction Hamiltonian and rotational Hamiltonian
respectively. The vibronic energy (in cm$^{-1}$) of a given electronic state
in an anharmonic oscillator approximation taking upto
first order term in $(v+\frac{1}{2})$ is,
\begin{eqnarray}
E_{Vib}(v)=(v+\frac{1}{2})\omega_e - (v+\frac{1}{2})^2 \omega_e\chi_e \label{E_vib}
\end{eqnarray}
where $\omega_e$ and $\omega_e\chi_e$ are harmonic vibrational frequency
and first correction due to anharmonicity respectively.
Now, for the case of spin-orbit interaction the orbital angular momentum ($\vec{L}$)
and spin angular momentum ($\vec{S}$) are strongly coupled to the internuclear axis.
If we denote the axial component of $\vec{L}$ and $\vec{S}$ as
$\Lambda$ and $\Sigma$ the spin-orbit coupling Hamiltonian will be,
\begin{eqnarray}
 H_{SO}= A_v \Lambda.\Sigma \label{H_so}
\end{eqnarray}

where $A_v$ is the spin-orbit coupling constant. $A_v$ depends
on the vibrational quantum number as per the following relation
derived by Brown and co-workers\cite{jmbrown_jmol-spec} (expanded upto the first order
of $(v+\frac{1}{2})$ term)
\begin{eqnarray}
 A_v = A_e \alpha_{A_e}(v+\frac{1}{2}) \label{A_v}
\end{eqnarray}

Therefore the spin-orbit Hamiltonian becomes,
\begin{eqnarray}
 H_{SO}= A_e \Lambda\Sigma - \alpha_{A_e}(v+\frac{1}{2}) \Lambda\Sigma \label{H_so}
\end{eqnarray}

In a molecular system rotation, vibration and electronic interactions influence one another.
For the ro-vibrational electronic spectra of a diatomic molecule,
the different angular momenta, i.e. electron spin angular
momentum ($\vec{S}$), electron orbital angular momentum ($\vec{L}$)
and angular momentum of nuclear rotation ($\vec{R}$)
can couple in various ways to form the resultant angular momentum $\vec{J}$.
These type of coupling are described by Hund’s coupling cases.
The ground state electronic multiplet, $X^2\Pi$, of SeD falls
into the category of Hund’s case (a) type of diatomic
molecule where electronic orbital angular momentum $\vec{L}$ is
weakly coupled with the nuclear rotation and strongly coupled
with the inter nuclear axis by electrostatic force i.e.
$|\frac{A_e}{B_e}|<<1$. Spin angular momentum ($\vec{S}$) is strongly
coupled to orbital angular momentum ($\vec{L}$) by spin orbit coupling. The electronic angular
momentum for a rotating diatomic molecule is defined as
$\Omega=\Lambda+\Sigma$ (Where $\Lambda$ and $\Sigma$
are the axial components of $\vec{L}$ and $\vec{S}$). Angular momentum of the
rotating nuclei ($\vec{R}$) is coupled to $\Omega$ to form total angular momentum $\vec{J}=\Omega+R$.
After neglecting the centrifugal force terms the rotational
energy in Hund’s case (a)\cite{bosser_JCPPC-biol}, rotational Hamiltonian will be
that of asymmetric top with $\Omega$ as the angular momentum about the
internuclear axis. The effective Hamiltonian in case for Hund’s case (a) is
\begin{eqnarray}
 H_{Rot} = |B_v| R^2 \label{H_rot}
\end{eqnarray}

where $B_v$ is the rotational constant. Hund’s case (a) is a good representation when $A_e\Lambda$ is much greater than
$B_v J$, where $A_e$ is the spin-orbit coupling constant. The rotational energy is given by,
\begin{eqnarray}
 E_{Rot}(J)= B_v [J(J+1)-\Omega^2]  \label{E_rot}
\end{eqnarray}

$B_v$ is dependent on the vibrational quantum number $v$. Expanding the rotational constant upto the first order of $(v+\frac{1}{2})$ term,
where $\alpha_e$ is rotation-vibration coupling constant
\begin{eqnarray}
 B_v=B_e-\alpha_e(v+\frac{1}{2})  \label{B_v}
\end{eqnarray}

Therefore the final form of the rotational energy can be written as,
\begin{eqnarray}
 E_{Rot}(J)&=& B_e[J(J+1)-\Omega^2] \nonumber \\
 &-&\alpha_e [J(J+1)-\Omega^2](v+\frac{1}{2}) \label{E_rot1}
\end{eqnarray}

Now, since the spin-orbit coupling and the vibronic energy are large
compared to the rotational energy, the total energy of the
Hamiltonian will clearly be the sum of the individual energies
and can be expanded in terms of spectroscopic constants by,
\begin{eqnarray}
 E(v,J)=&\pm&[A_e - \alpha_{A_e}(v+\frac{1}{2})] + (v+\frac{1}{2})\omega_e \nonumber \\
         &-& (v+\frac{1}{2})^2\omega_e\chi_e + B_e[J(J+1)-\Omega^2] \nonumber \\
         &-&\alpha_e [J(J+1)-\Omega^2](v+\frac{1}{2}) \label{E_vj}.
\end{eqnarray}

Here the top and bottom sign denotes the $^2\Pi_{\frac{3}{2}}$ and $^2\Pi_{\frac{1}{2}}$
state respectively. $\alpha_{A_e}$ may be represented as the difference of the
harmonic frequencies of $^2\Pi_{\frac{3}{2}}$ and $^2\Pi_{\frac{1}{2}}$
state when considered independently\cite{jmbrown_jmol-spec}. The total energy can be separated into the
summation of $J$ dependent and the $J$ independent part,
\begin{eqnarray}
 E(v,J)= G(v)+F_v(J)
\end{eqnarray}
where,
\begin{eqnarray}
 G(v)&=& \pm\frac{1}{2}A_e - B_e\Omega^2 + (\omega_e\mp \frac{1}{2}\alpha_{A_e} \\ \nonumber
 &+&\alpha_{A_e}\Omega^2 )(v+\frac{1}{2})\\ \nonumber
       &-& (v+\frac{1}{2})^2\omega_e\chi_e \\
 F_v(J)&=& B_e[J(J+1)] -\alpha_e [J(J+1)](v+\frac{1}{2})
\end{eqnarray}

Separating the energy in terms of $J$ dependency allows us to pick up vibrational levels belonging
to different spin-orbit coupled electronic states having similar energy.
As a further refinement to the energy level difference, $J$'s belonging to same or different
vibrational levels can be chosen as per the experimental requirements.

\section{COMPUTATIONAL DETAILS}

The ground electronic state $X^2\Pi$ of Selenium Deuteride radical is thoroughly studied in this paper.
The ground state electronic configuration of $X^2\Pi$ SeD is
$(3s_{Se})^2 (3p_{z Se}+1s_H)^2 (3p_{\pi Se})^2 (3p_{\pi Se})^1$.
If one electron is transferred from $(3p_{z Se}+1s_H)$ orbital to $(3p_{\pi Se})$
the first excited state $^2\Sigma^+$ is formed with the configuration
$(3s_{Se})^2 (3p_{z Se}+1s_H)^2 (3p_{\pi Se})^2 (3p_{\pi Se})^2$.
The theoretical characterization of potential energy curves
for the ground state and the first excited state over an
extended inter nuclear separation until dissociation
requires the account of both static and dynamic correlation effects.
In this study for the ground state, these effects has been
included by the use of state-averaged complete active space SCF(SA-CASSCF)\cite{warner_jcp,knowles_cpl85}
calculation on each doublet spin symmetry
followed by single and double electron excitation on top of the
zeroth order multi reference wavefunction (MRCISD).
SA-CASSCF step involved the two states of symmetry
$B_1(\Pi_x)$ and $B_2(\Pi_y)$ in the $C_{2V}$ point group representation,
the symmetries in the parenthesis are the corresponding one in the
$C_{\infty V}$ point group. The active space consists the distribution of 7 electrons over 5
orbitals(CAS(7,5)). Since single and double electron excitation on top of CAS wave
function is computationally very demanding, so Configuration State Function(CSF) with coefficients greater than
0.01 only are included to construct the new zero-order CI space.
A further reduction of the dimension of the CI wavefunction has been made with internally
contracted configuration interaction (IC-MRCISD) approach\cite{ knowles_cpl88,warner-jcp}
by restricting the core occupation to $(7A_1,3B_1,3B_2,1A_2)$
where A and B denotes the symmetries of the irreducible representation in
$C_{2V}$ point group symmetry.

The molecular orbitals constructing the CSF's are the natural orbitals which are
obtained by diagonalization of state-averaged ($B_1$ and $B_2$ state in
$C_{2V}$ point group symmetry) density matrix.
Douglas-Kroll contracted correlation consistent Dunning's
VnZ-DK(n=3-5)\cite{dunning_jcp,wilson_jcp} basis sets, employed for both the atoms,
used in the expansion of natural orbital. The second-order Douglas-Kroll-Hess
Hamiltonian has been used for all MRCI and CASSCF computations to account
for the scalar relativistic effects\cite{douglas_ap,hess_pra}.

In the next step, the major focus is to determine the spin-orbit coupling.
Technically spin-orbit contribution is computed
using two steps: first, the SO Hamiltonian is added in a fashion of
general first order perturbation procedure to the electronic
Hamiltonian to construct the total Hamiltonian of the form
$\hat{H}=\hat{H}_{el}+\hat{H}_{SO}$. The spin-orbit matrix elements
$\hat{H}_{SO}$ are calculated between the internal configurations
(i.e. no electron in the external orbitals) $^2\Pi(^2B_1)$ and $^2\Pi(^2B_2)$
with the spin-orbit full Breit-Pauli(BP)\cite{langhoff_jcp} operator of the form
\begin{eqnarray}
 \hat{H}_{SO}=\frac{1}{2m^2c^2}\Bigg[&\sum_{i}&\sum_{\alpha}\frac{Z_{\alpha}e^2}{r_{i\alpha}^3}\hat{I}_{i\alpha}.\hat{S}_{i} \nonumber \\
 &-&\sum_{i}\sum_{\alpha}\frac{e^2}{r_{ij}^3}\hat{I}_{ij}.(\hat{S}_i+2\hat{S}_j)\Bigg]
\end{eqnarray}

which contains both one and two electron terms. Here $\hat{I}$ and $\hat{S}$ are orbital and spin angular momentum operators,
$i$ and $\alpha$ denotes electron and nucleus respectively.

In the next step, $\hat{H}=\hat{H}_{el}+\hat{H}_{SO}$ matrix is diagonalised in the basis of SA-CASSCF/IC-MRCISD(7,5) wavefunctions to yield the
desired spin-orbit splitting directly.

In order to improve the level of description this spin-orbit splitting is added as $\it a$ $\it posteriori$ correction to the corresponding
MRCI+Q energy at each internuclear separation, where +Q denotes the quadruple excitation
corrected by Davidson’s method\cite{langhoff_ijqc,rawlings_ijqc}.These calculations have been carried out with the MOLPRO\cite{molpro} suite
of programs.

\section{The potential energy function}

Among the functions that was proposed to fit the Analytical Potential Energy Functions(APEF) of diatomic
molecules, Murrel Sorbie (MS) potential energy function
seems to be the best one\cite{yang_j-mol-str-2003,yang_j-mol-phys,yang_j-mol-str-2001,Zhu-Sc-press,yang_j-mol-str-2004-1,yang_j-mol-str-2004-2}.
The interaction potential energies of many neutral and cationic diatomic
molecules can be accurately reproduced by this function and has been used to deduce APEFs for many
molecules\cite{yang_j-mol-str-2003,yang_j-mol-phys,yang_j-mol-str-2001,Zhu-Sc-press,yang_j-mol-str-2004-1,yang_j-mol-str-2004-2}.
The general form of MS function is given by\cite{murrel_sorbie_far-trans}

\begin{equation}
   V(\rho)=-D_e(1+\sum_{i=1}^na_i\rho^i)exp(-a_1\rho)
\end{equation}

Usually, satisfactory results can be obtained when
$n$ equals 3. In order to get accurate data, the following form of MS function is used\cite{murrel_sorbie_far-trans,murrel_sorbie_mol-phys}
\begin{equation}
   V(\rho)=-D_e(1+a_1\rho+a_2\rho^2+a_3\rho^3)exp(-a_1\rho)
\end{equation}

where $\rho=R-R_{e}$ is the inter atomic distance, $R_e$ is the equilibrium distance
and $D_e$ is the dissociation energy.
The quadratic($f_2$), cubic($f_3$) and quatric($f_4$) force constants can be derived
by the M-S function and then spectroscopic parameters harmonic frequency
($\omega_e$) , anharmonicity factor ($\omega_e\chi_e$) rotational constant($B_e$)
and vibration-rotation coupling constant ($\alpha_e$) can be calculated by the following relations,
\begin{eqnarray}
  f_2&=&D_e(a_1^2-2a_2) \\
  f_3&=&6D_e(a_1a_2-a_3-\frac{a_1^3}{3}) \\
  f_4&=&D_ea_1^4-6f_2a_1^2-4f_3a_1 \\
  B_e&=&\frac{h}{8\pi^2c\mu R_e^2} \\
  \omega_e&=&\sqrt{\frac{f_2}{4\pi^2mc^2}} \\
  \alpha_e&=&-\frac{6B_e^2}{\omega_e}(\frac{f_3R_e}{3f_2}+1) \\
  \omega_e\chi_e&=&\frac{B_e}{8}\Bigg[-\frac{f_4R_e^2}{f_2}+15(1+\frac{\omega_e\alpha_e}{6B_e^2})^2\Bigg].
\end{eqnarray}

Once these parameters are obtained from the calculation, it is straightforward to calculate the amplification factor for our
purpose.

\section{RESULT AND DISCUSSION}

\begin{table}[htbp]
 \begin{center}
 \caption{\label{tab1}Spectroscopic parameters derived for the most abundant isotope $^{80}SeD(\mu=1.9645891 a.u)$
 from Murrel-Sorbie curve fitting with different Dunning’s basis set with the Spin-Orbit coupling at equilibrium.}
\resizebox{0.5\textwidth}{!}{
   \begin{tabular}{|l|c|c|c|c|c|c|c|}
   \hline
   Basis & Electronic & $R_e$   &   $\omega_e$ &      $\omega_e\chi_e$     &    $B_e$    &   $\alpha_e$   &   $A_e$    \\
         & State      & $\AA{}$ &   $cm^{-1}$  &  $cm^{-1}$  &  $cm^{-1}$  &  $cm^{-1}$  & $cm^{-1}$  \\
   \hline \hline
   a-VTZ-DK & $X^2\Pi$              & 1.4711 & 1754.20 & 22.598 & 3.9621  & 0.07327  &  -1759.72  \\
            & $^2\Pi_{\frac{3}{2}}$ & 1.4711 & 1750.08 & 22.429 & 3.9621  & 0.07360  &   \\
            & $^2\Pi_{\frac{1}{2}}$ & 1.4711 & 1758.64 & 22.783 & 3.9621  & 0.07294  &   \\
%   \hline
   VQZ-DK   &  $X^2\Pi$             & 1.4689 &  1769.06  & 22.576 & 3.9735 & 0.0724  &  -1768.01   \\
            & $^2\Pi_{\frac{3}{2}}$ & 1.4689 &  1764.85  & 22.402 & 3.9735 & 0.0727  &    \\
            & $^2\Pi_{\frac{1}{2}}$ & 1.4689 &  1773.54  & 22.765 & 3.9735 & 0.0721  &    \\
%   \hline
   a-VQZ-DK &  $X^2\Pi$             & 1.4689 &  1764.39  & 22.588 & 3.9735 & 0.0732  &  -1767.81   \\
            & $^2\Pi_{\frac{3}{2}}$ & 1.4689 &  1760.41  & 22.426 & 3.9735 & 0.0735  &    \\
            & $^2\Pi_{\frac{1}{2}}$ & 1.4689 &  1768.66  & 22.762 & 3.9735 & 0.0728  &    \\
%   \hline
   V5Z-DK   &  $X^2\Pi$             & 1.4689 &  1766.88  & 22.441 & 3.9735 & 0.0724  &  -1770.26   \\
            & $^2\Pi_{\frac{3}{2}}$ & 1.4689 &  1762.79  & 22.273 & 3.9735 & 0.0728  &    \\
            & $^2\Pi_{\frac{1}{2}}$ & 1.4689 &  1771.27  & 22.623 & 3.9735 & 0.0721  &    \\
%   \hline
   a-V5Z-DK   &  $X^2\Pi$             & 1.4689 &  1765.11  & 22.442 & 3.9735 & 0.0727  &  -1770.26   \\
            & $^2\Pi_{\frac{3}{2}}$ & 1.4689 &  1761.11  & 22.280 & 3.9735 & 0.0730  &    \\
            & $^2\Pi_{\frac{1}{2}}$ & 1.4689 &  1769.41  & 22.618 & 3.9735 & 0.0724  &    \\
   \hline \hline
         & Electronic & $R_e$     &   $\nu_0$\footnote[1]{$\nu_0$ is the fundamental transition frequency.$\nu_0$ is defined as $\nu_0=\omega_e-2\omega_e\chi_e$} &      $\omega_e\chi_e$     &    $B_e$    &   $\alpha_e$   &   $A_e$    \\
         & State      & $\AA{}$   &   $cm^{-1}$  &  $cm^{-1}$  &  $cm^{-1}$  &  $cm^{-1}$  & $cm^{-1}$  \\
   \hline
   Expt.    &  $X^2\Pi$             & 1.4640 &  1677.05  & 21.35   & 4.00310 & 0.07985  &  -1762.696   \\
            & $^2\Pi_{\frac{3}{2}}$ &   -    &     -     &    -    &    -    &     -    &    \\
            & $^2\Pi_{\frac{1}{2}}$ &   -    &     -     &    -    &    -    &     -    &    \\
%            &\hspace{10mm}          &\hspace{10mm}&\hspace{10mm}&\hspace{10mm}&\hspace{10mm}&\hspace{10mm}&\hspace{10mm} \\

   \hline \hline
   \end{tabular}
  }
  \end{center}
 \end{table}

\begin{figure}[h]
 %\hglue -1.5cm
 \includegraphics[angle=-90,scale=0.35]{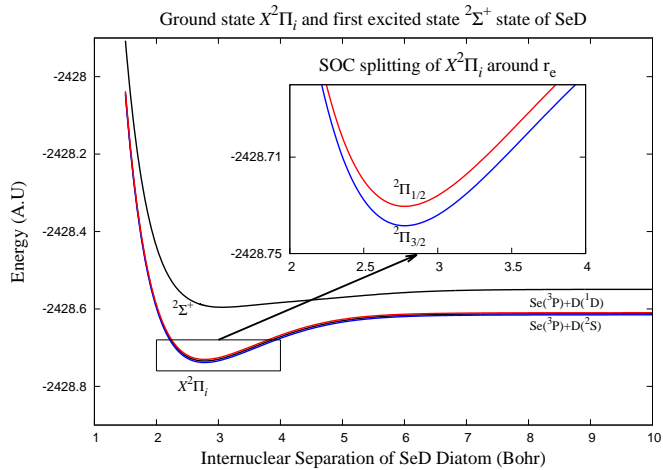}\\
 \caption{\label{fig1} The PESs of the ground state $X^2\Pi$ and the first excited state $A^2\Sigma^+$
           of SeD radical and partly magnified PESs of the states $^2\Pi_{\frac{3}{2}}$ and $^2\Pi_{\frac{1}{2}}$
           near equilibrium(inset) at the MRCI+Q/a-V5Z-DK level of theory.}
 \end{figure}

Three lowest states of Selenium Deuteride radical $X^2\Pi,^2\Pi_{\frac{3}{2}}$ and $^2\Pi_{\frac{1}{2}}$
are least square fitted to the Murrel-Sorbie function to get the parameters
$a_1,a_2,a_3,R_e$ and $D_e$ of the corresponding states. By using the parameters into the respective equations the
spectroscopic parameters are evaluated for the most abundant isotopes of Se i.e. $^{80}$Se for the SeD molecule with different
correlation consistent with Dunning’s basis sets. The potential energy surfaces at the MRCI+Q/a-V5Z-DK for the states
$X^2\Pi,^2\Pi_{\frac{3}{2}}$ and $^2\Pi_{\frac{1}{2}}$ of SeD are shown in Fig.~\ref{fig1}.
The plotted potential energy curves are smooth and show no presence of unphysical kinks along the whole surface.

\begin{figure}[h]
% \hglue -1.5cm
 \includegraphics[angle=-90,scale=0.35]{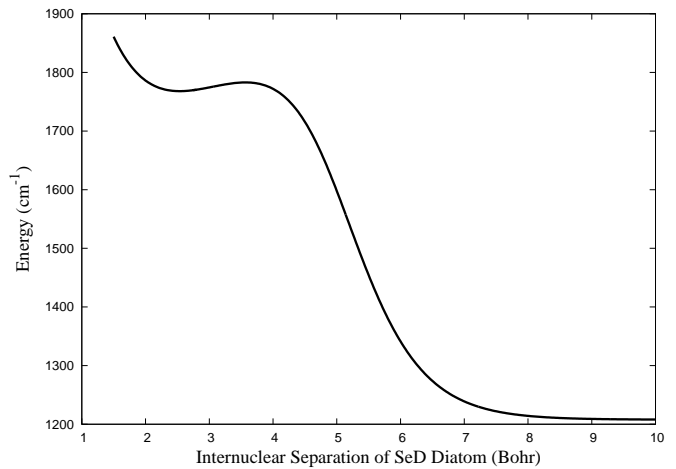}\\
 \caption{\label{fig2} The curve for the vertical transition energy of SeD $X^2\Pi_{\frac{3}{2}\rightarrow \frac{1}{2}}$
          transition energy($cm^{-1}$) vis the internuclear separation at MRCI+Q/a-V5Z-DK level of theory .}
 \end{figure}

The variation of spin-orbit energy difference between the two spin orbit components
$X^2\Pi_{\frac{3}{2}}$ and $^2\Pi_{\frac{1}{2}}$ with the change in
internuclear separation between Se and D is presented in Fig.~\ref{fig2}.
This energy interval can be regarded as the vertical transition energy from
$X^2\Pi_{\frac{3}{2}}$ to $^2\Pi_{\frac{1}{2}}$ which is  determinded by the SOC splitting of the ground electronic state
$X^2\Pi$. Increasing the inter-atomic separation from the equilibrium
distance 2.776 a.u, the curve exhibits a little increase(~$12cm^{-1}$) upto 3.56 a.u and then a gradual decrease is observed to about
1207.91$cm^{-1}$ at 10 a.u. The curve shows the SOC between $X^2\Pi_{\frac{3}{2}}$ and $^2\Pi_{\frac{1}{2}}$ is lower than 1780
$cm^{-1}$ at all inter nuclear separation which supports the perturbative treatment of SOC in this case.

Relevant spectroscopic parameters such as, harmonic frequency $\omega_e$,
anharmonicity factor ($\omega_e\chi_e$), rotational constant($B_e$)
and vibration-rotation coupling constant ($\alpha_e$) along with equilibrium
bond length and Spin-Orbit Coupling(SOC) are tabulated with
different electron correlation consistent Dunnings basis set in Tab.~\ref{tab1}.

It is evident from Table~\ref{tab1} that the predicted equilibrium bond lengths of SeD using MRCI at different basis sets reveal almost
no discernible variation from quadruple zeta quality basis sets to quintuple zeta basis sets and is less than
$0.0001\AA{}$. The estimated $R_e$ is also in excellent agreement with the previously measured bond lengths of SeD
($1.4640\AA{}$) from Laser Magnetic Resonance based experiments\cite{jmbrown_jmol-spec}. We find the deviation at MRCI/a-V5Z-DK is
less than $0.005\AA{}$.
Moreover, the estimated spectroscopic constants, also show excellent convergence at better quality basis sets.
In general theoretically estimated spectroscopic constants also exhibit excelent agrement with the experimentally determined values.
However for $\omega_e$, we find that the agrement between the experimental and theoretical value can be termed satisfactory at best.
Decades back Brown and co-workers have determined the band origin of SeD vibrationional spectrum to be 1677.05 $cm^{-1}$ \cite{jmbrown_jmol-spec}.
Using the equation $\nu_0=\omega_e-2\omega_{e}\chi_{e}$, where $\nu_0$ is the fundamental vibrational band origin the $\omega_e$
is estimated to be 1719.75 $cm^{-1}$ \cite{jmbrown_jmol-spec}. On the contrary the $\omega_e$ predicted from the Murrel-Sorbie fit to the MRCI
potential energy surface predicts a $\omega_e$ of 1765.11 $cm^{-1}$.
Harmonic frequency derived numerically by determining the 2nd derivative of energy
with respect to nuclei displacement about the equilibrium geometry is predicted to be 1712.12 $cm^{-1}$.
We have checked also the $\omega_e$ with other reliable electronic structure methods such as coupled cluster singles and doubles
(CCSD)\cite{ccsd-1,ccsd-2,ccsd-3}
and coupled cluster singles and doubles with perturbative Triples corrections (CCSD(T))\cite{ccsdt-1} with electron correlation consistent dunning's basis sets.
These methods can provide accurate estimates for harmonic frequencies($\omega_e$) as at equilibrium bond length if the system under consideration
can be well approximated through a single determinant wave-function. All of these values for harmonic frequencies are reported in Table~\ref{tab2}.
The amplification factor, K is dependent on the harmonic frequency. Hence using different harmonic frequencies we get a wide ranging spread
for the amplification factor as shown in Table~\ref{tab2}. The amplification factor can be as high as 1070.
If use only theoretical estimates the maximum amplification factor can be 350. On the lower side it can be 34.
Even if we consider the lowest amplification factor one can safely conclude that this molecule can be an effective probe
for measuring space time variation of fundamental constants.

\begin{table}[htbp]
 \begin{center}
\caption{\label{tab2} Amplification factor(K) predicted at different level of theory with the experimental SOC}
\resizebox{0.5\textwidth}{!}{
 \begin{tabular}{|l|c|c|c|c|c|}
   \hline
   Isotope    & Electronic &  Level of Theory  &  $\omega_e$\footnote{harmonic frequency($\omega_e$) obtained numerically}  &  $A_e$ or $\omega_f$\footnote{Expt. value\cite{jmbrown_jmol-spec}}  &  Amplification factor(K)  \\
              &   State    &                   &  $cm^{-1}$   &  $cm^{-1}$                                  &   $K=\frac{\omega_f}{\omega_f - v \omega_e}$              \\
   \hline \hline
   $^{80}SeD$ &  $^2\Pi$   & CCSD/a-VQZ-DK & 1761.04  &   -1762.696  &    1064.42 \\
   \hline
   $^{80}SeD$ &  $^2\Pi$ &    CCSD(T)/a-VQZ-DK     & 1738.05   &  -1762.696  &    71.52 \\
   \hline
   $^{80}SeD$ &   $^2\Pi$ &  MRCI/a-VQZ-DK       & 1712.12   &  -1762.696  &    34.85 \\
   \hline
   $^{80}SeD$ &    $^2\Pi$ &   Expt.\cite{jmbrown_jmol-spec}      &  1719.75   &  -1762.696  &   41.04 \\
 %             & &(from Murrel-Sorbie fit of full PES &  &  & \\
   \hline \hline
   \end{tabular}
   }
 \end{center}
 \end{table}

Since from the above table, the spin-orbit coupling constant $A_e$ and harmonic frequency
$\omega_e$ are very similar in magnitude for SeD radical in its ground state electronic multiplet
$X^2\Pi_i$, $G(v+1)^{\frac{3}{2}}$ is quasi-degenerate with
$G(v)^{\frac{1}{2}}$ level for $v=0,1,2,3,...$. As mentioned in the introduction,
for measuring space-time variation of fundamental physical constants, we have to
have a large value of amplification factor ($K$) for the transition between
quasi-degenerate vibronic levels. The energy difference
between the quasi-degenerate vibronic levels can be expressed as,
\begin{eqnarray}
 \varDelta G(v)&\equiv& G(v)^{\frac{1}{2}} - G(v+1)^{\frac{3}{2}} , \nonumber \\
 &=& -A_e-\omega_e+2B_e-2\alpha_e\\ \nonumber &+&(2\omega_e\chi_e+\alpha_{A_e})(v+1)
\end{eqnarray}

For the most abundant four isotopes of SeD at the MRCI+Q/a-V5Z-DK
level of theory, the change of $\varDelta G(v)$ is tabulated in Table~\ref{tab3}.
$\varDelta G(v)$ is positive for all vibrational levels and increases with
the vibrational quantum number due to anharmonicity. For $v=0$ the
vibrational levels of the two states become closest to each other.

 \begin{table}[htbp]
 \begin{center}
 \caption{\label{tab3} Difference between quasi-degenerate vibronic states for the four
 most abundant isotope SeD with increasing vibrational quantum number.}
\resizebox{0.5\textwidth}{!}{
 \begin{tabular}{|l|c|c|c|c|}
   \hline
   Isotope    &  $\varDelta G(v)$  &  $\varDelta G(v=0)$  &  $\varDelta G(v=1)$  &  $\varDelta G(v=2)$  \\
              &  $cm^{-1}$         &  $cm^{-1}$           &  $cm^{-1}$           &  $cm^{-1}$           \\
   \hline \hline
   $^{77}SeD$ &  11.85+36.62($v+1$) &  48.47   &   85.09 &  121.71 \\
   \hline
   $^{79}SeD$ &  12.54+36.59($v+1$) &  49.13   &   85.72 &  122.31 \\
   \hline
   $^{80}SeD$ &  12.84+35.58($v+1$) &  49.40   &   83.98 &  119.57 \\
   \hline
   $^{82}SeD$ &  13.34+35.58($v+1$) &  49.89   &   86.44 &  122.99 \\
   \hline \hline
   \end{tabular}
   }
 \end{center}
 \end{table}

Since for $v=0$ the vibrational levels of the two states come within a
$50$ $cm^{-1}$, rotational states with $J$-value from the two states
interact significantly with each other. Rotational energies of both
the doublet states are expressed with the same expression i.e.
$F_v(J)^{(\frac{1}{2})}=F_v(J)^{(\frac{3}{2})}=B_e[J(J+1)-\alpha_e J(J+1)(v+\frac{1}{2})]$.
Therefore the energy associated with the microwave transition is
$\varDelta F_v(J)=F_v(J)^{(\frac{1}{2})}-F_v(J)^{(\frac{3}{2})}$.
Now the selection rule for microwave transition is
$\varDelta F_v(J)=\pm 1$ i.e. $\varDelta L+\varDelta J=\pm 1$,
which leads to two possibility, one is $\varDelta L =\pm 1,\varDelta J=0$ (only observed
for open shell molecule which leads to Q-Branch spectra)
and another is $\varDelta L =0,\varDelta J=\pm 1$ (Which leads to P and R-Branch spectra).
So the overall selection rule for this kind of doublet species is
$\varDelta J=0,\pm 1$. For $\varDelta J=0$ transitions, there are no change in the rotational energy.
So we are considering only those transitions which follow the selection rule
$\varDelta J=\pm 1$. For $\varDelta J=+1$, $\varDelta F_{v=1}(J)=2B_e(J+1)-\alpha_eJ(J+1)(J+3)$
and for $\varDelta J=-1$, $\varDelta F_{v=1}(J)= -2B_eJ-\alpha_eJ(J-2)$.

Now the transitions of interest are those which lead to $\varDelta E(v,J)\approx 0$ i.e.
$\varDelta G(v)+\varDelta F_v(J)\approx 0$. Since for $v=1$ vibronic levels of
the two doublet states come closest and $\varDelta G(v=0)$ is a positive quantity
$\varDelta J$ have to be -1. Therefore $\varDelta G(v=0) + \varDelta F_v(J)\approx 0$.
\begin{eqnarray}
 \varDelta G(v=0)&=&-\varDelta F_v(J) , \nonumber \\
 &=&-(-2B_eJ-\alpha_eJ(J-2)) , \nonumber \\
 &=&2B_eJ+\alpha_eJ(J-2)
\end{eqnarray}

From the table of the spectroscopic parameters we notice the fact that
$\alpha_e << B_e$ in magnitude and we neglect the term containing
$\alpha_e$ to convert the equation into linear equation
$\varDelta G(v=0)=2B_eJ$ and solve for $J$ leading to
\begin{eqnarray}
 J=\frac{\varDelta G(v=0)}{2B_e}=6.21\approx 6 \nonumber
\end{eqnarray}

For open shell systems $J$ is essentially half integer, the two appropriate choices of
$J$ be $J=6\frac{1}{2}$ and $J=5\frac{1}{2}$.

\subsection{Variation of ro-vibronic transition frequency with respect to variations of $\alpha$ and $\mu$:}

\begin{figure}
\begin{center}
                \includegraphics[angle=0,width=0.5\textwidth]{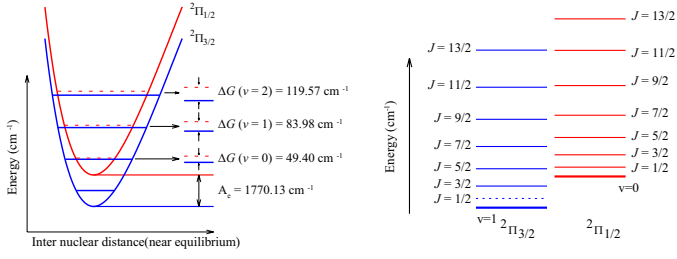}
               \caption{(left) The magnitude of Vibronic levels of the two doublet states of $^{80}SeD$. The blue lines represents
          $^2\Pi_{\frac{3}{2}}$ and red line represents $^2\Pi_{\frac{1}{2}}$ MRCI+Q/a-V5Z-DK level of theory. (right) Rotational levels of the two doublet states. The blue (solid) lines represents $^2\Pi_{\frac{3}{2}}$
          and red (dashed) line represents $^2\Pi_{\frac{1}{2}}$. Since $J$ is always $\geq\Omega$ so for $^2\Pi_{\frac{3}{2}}$,
          $J=\frac{1}{2}$ rotational level is not observed.}
          \label{fig34}
\end{center}
\end{figure}

The ro-vibrational energy difference between two electronic states can be expressed as,
\begin{eqnarray}
 \varDelta E_v(J)&=&\varDelta G(v)+\varDelta F_v(J) , \nonumber \\
  &=& A_e-\omega_e-\alpha_e+2B_e+\\\nonumber
  & &v(2\omega_e\chi_e-\alpha_{A_e})-2B_eJ-\alpha_eJ(J-2)
\end{eqnarray}
So for the variation in energy difference $\varDelta E_v(J)$ in terms of variation in $\alpha$ and $\mu$ can be expressed as,
\begin{eqnarray}
 \delta E_v(J)&=&\delta(A_e-\omega_e-\alpha_e+2B_e+\\ \nonumber
  & &v(2\omega_e\chi_e-\alpha_{A_e})-2B_eJ-\alpha_eJ(J-2)) , \nonumber \\
 &=&\delta (A_e-\omega_e)
\end{eqnarray}
As other terms are negligible compared to $A_e$ and $\omega_e$ they are neglected in equation (30). Now the spin-orbit constant
$A_e$ varies as $\sim Z^2 \alpha^2 E_H$ and $\omega$ varies as
$M_r^{-\frac{1}{2}}\mu^{-\frac{1}{2}}E_H$ as mentioned in the introduction
so overall variation in transition energy will be,
\begin{eqnarray}
 \delta E_v(J)&=&\delta (A_e-\omega_e) , \nonumber \\
 &\cong& 2A_e (\frac{\delta \alpha}{\alpha} + \frac{1}{4} \frac{\delta \mu}{\mu}) , \nonumber \\
 &=& 3540 cm^{-1} (\frac{\delta \alpha}{\alpha} + \frac{1}{4} \frac{\delta \mu}{\mu})
\end{eqnarray}
For SeD the values are shown in Fig.~\ref{fig34}. Therefore a large enhancement factor may be obtained by proper choice of a molecular probe,
in this case the SeD radical which is likely to be found under certain
astrophysical conditions like an AGB star.

\section{CONCLUSION}

To summarize, for SeD molecule, we have analyzed the sensitivity of the
ro-vibronic spectrum to variations in the fundamental
physical constants. We have found enhanced sensitivity for a
number of low frequency microwave transition within
$^2\Pi_{\frac{1}{2}}(v=0)$ and $^2\Pi_{\frac{3}{2}}(v=1)$
which may enhance the amplification factor upto the order of $\sim 350$.
We acknowledge the fact that the data produced in the calculation should
not be considered as accurate as microwave frequency because the error bar of the
MRCISD+Q level of theory can be as large as $\sim 100 cm^{-1}$.
Fairly accurate data can only be obtained from high precision
laboratory experiments. So, experimental evidence is necessary
on the molecule to confirm our findings.

\section{Acknowledgments}

G.G and A.S acknowledge the Council of Scientific and Industrial research (CSIR) India for a their Research Fellowship. A.P would like
to thank CEFIPRA (Project No. IFC/4705-3/2012/3025) for financial support.

\end{document}